\documentclass[twocolumn,english,aps,preprintnumbers,amsmath,amssymb,superscriptaddress]{revtex4}
\usepackage[latin9]{inputenc}
\usepackage{amsmath}
\usepackage{graphicx}

\makeatletter
\@ifundefined{textcolor}{}
{%
 \definecolor{BLACK}{gray}{0}
 \definecolor{WHITE}{gray}{1}
 \definecolor{RED}{rgb}{1,0,0}
 \definecolor{GREEN}{rgb}{0,1,0}
 \definecolor{BLUE}{rgb}{0,0,1}
 \definecolor{CYAN}{cmyk}{1,0,0,0}
 \definecolor{MAGENTA}{cmyk}{0,1,0,0}
 \definecolor{YELLOW}{cmyk}{0,0,1,0}
}

\usepackage{dcolumn}
\usepackage{bm}
\usepackage{color}
\usepackage[final]{pdfpages}

\makeatother

\usepackage{babel}

\begin{document}

\preprint{preprint(\today)}

\title{Local spectroscopic evidence for a nodeless magnetic kagome superconductor CeRu$_{2}$}

\author{C.~Mielke III$^{\dag}$}
\email{charles-hillis.mielke-iii@psi.ch} 
\affiliation{Laboratory for Muon Spin Spectroscopy, Paul Scherrer Institute, CH-5232 Villigen PSI, Switzerland}
\affiliation{Physik-Institut, Universit\"{a}t Z\"{u}rich, Winterthurerstrasse 190, CH-8057 Z\"{u}rich, Switzerland}

\author{H.~Liu}
\thanks{These authors contributed equally to the paper.}
\affiliation{Beijing National Laboratory for Condensed Matter Physics and Institute of Physics, Chinese
Academy of Sciences, Beijing 100190, China.}
\affiliation{University of Chinese Academy of Sciences, Beijing 100049, China.}

\author{D.~Das}
\affiliation{Laboratory for Muon Spin Spectroscopy, Paul Scherrer Institute, CH-5232
Villigen PSI, Switzerland}

\author{J.-X.~Yin}
\affiliation{Laboratory for Topological Quantum Matter and Advanced Spectroscopy (B7), Department of Physics,
Princeton University, Princeton, New Jersey 08544, USA}

\author{L.Z. Deng}
\affiliation{Department of Physics and Texas Center for Superconductivity, University of Houston, Houston, TX}

\author{J. Spring}
\affiliation{Physik-Institut, Universit\"{a}t Z\"{u}rich, Winterthurerstrasse 190, CH-8057 Z\"{u}rich, Switzerland}

\author{R.~Gupta}
\affiliation{Laboratory for Muon Spin Spectroscopy, Paul Scherrer Institute, CH-5232
Villigen PSI, Switzerland}

\author{M.~Medarde}
\affiliation{Laboratory for Multiscale Materials Experiments, Paul Scherrer Institut, CH-5232 Villigen PSI, Switzerland}

\author{C.-W. Chu}
\affiliation{Department of Physics and Texas Center for Superconductivity, University of Houston, Houston, TX}
\affiliation{Lawrence Berkeley National Laboratory, Berkeley, California 94720, USA}

\author{R.~Khasanov}
\affiliation{Laboratory for Muon Spin Spectroscopy, Paul Scherrer Institute, CH-5232 Villigen PSI, Switzerland}

\author{Z.M. Hasan}
\affiliation{Laboratory for Topological Quantum Matter and Advanced Spectroscopy (B7), Department of Physics, Princeton University, Princeton, New Jersey 08544, USA}
\affiliation{Princeton Institute for the Science and Technology of Materials, Princeton University, Princeton, New Jersey 08540, USA}
\affiliation{Materials Sciences Division, Lawrence Berkeley National Laboratory, Berkeley, California 94720, USA}
\affiliation{Quantum Science Center, Oak Ridge, Tennessee 37831, USA}

\author{Y.~Shi}
\affiliation{Beijing National Laboratory for Condensed Matter Physics and Institute of Physics, Chinese
Academy of Sciences, Beijing 100190, China.}
\affiliation{University of Chinese Academy of Sciences, Beijing 100049, China.}

\author{H.~Luetkens}
\affiliation{Laboratory for Muon Spin Spectroscopy, Paul Scherrer Institute, CH-5232 Villigen PSI, Switzerland}

\author{Z.~Guguchia}
\email{zurab.guguchia@psi.ch} 
\affiliation{Laboratory for Muon Spin Spectroscopy, Paul Scherrer Institute, CH-5232 Villigen PSI, Switzerland}

\begin{abstract}

We report muon spin rotation (${\mu}$SR) experiments on the microscopic properties of superconductivity and magnetism in the kagome superconductor CeRu$_{2}$ with $T_{\rm c}$~${\simeq}$~5~K. From the measurements of the temperature-dependent magnetic penetration depth ${\lambda}$, the superconducting order parameter exhibits nodeless pairing, which fits best to an anisotropic $s$-wave gap symmetry. We further show that the $T_{\rm c}$/$\lambda^{-2}$ ratio  is comparable to that of  unconventional superconductors. Furthermore, the powerful combination of zero-field (ZF)-${\mu}$SR and high-field ${\mu}$SR has been used to uncover magnetic responses across three characteristic temperatures, identified as $T_1^*$~${\simeq}$~110~K, $T_2^*$~${\simeq}$~65~K, and $T_3^*$~${\simeq}$~40~K. 
Our experiments classify CeRu$_{2}$ as an exceedingly rare nodeless magnetic kagome superconductor.

\end{abstract}

\maketitle

\section{Introduction}

The unique kagome lattice, formed by an interwoven network of corner-sharing triangles, is well-known to host many fascinating physical phenomena \cite{GuguchiaCSS,JXYin2,MielkeKVS,JiangpingHu,TNeupert,Yu,Ortiz}. Arising from the natural geometrical frustration, band structure calculations reveal several characteristic features in this atomic lattice, most frequently flat bands, van Hove singularities, and Dirac nodes, which can influence the electronic properties and give rise to topologically nontrivial phases when found near the Fermi energy. One of the most rare phenomena exhibited by kagome lattice materials is superconductivity, which often hosts competing magnetic \cite{MielkeKVS} or otherwise unconventional \cite{MielkeLRS} features. In our recent work on LaRu$_{3}$Si$_{2}$, which we identified as a robust $s$-wave kagome superconductor, we found that the critical temperature cannot be fully explained by electron-phonon coupling, but experiences additional enhancement from typical kagome band structure features found near the Fermi energy \cite{MielkeLRS}. In further explorations of the recently discovered KV$_{3}$Sb$_{5}$ \cite{BOrtiz3,YJiang} and sister compounds \cite{MielkeKVS,GuguchiaRVS,NShumiya,Ritu,Ortiz}, we have identified time-reversal symmetry-breaking associated with the charge ordering transition at $T_{\rm co}$~${\simeq}$~80~K, two orders of magnitude higher than the superconducting transition, $T_{\rm c}$~${\simeq}$~1.1~K.

While the distorted Laves-phase superconductor CeRu$_{2}$ takes a cubic structure (Fig. 1a) \cite{HuxleyPolN} with two different Ce sites, it reveals a pristine Ru kagome lattice (Fig.~1b) that contributes to the electronic properties. Indeed, the normal state band structure features a kagome flat band, Dirac points and van Hove singularities formed by the Ru-$dz^{2}$ orbitals near the Fermi level \cite{Deng}, which are predicted to support topologically nontrivial states \cite{Deng}. Photoemission studies show the highly itinerant nature of the Ce electrons in CeRu$_{2}$, identifying a strong hybridization effect in the itinerant 4$f$-bands \cite{Sekiyama,JSKang}. Additionally, much attention has been given to the unusual superconducting state in CeRu$_{2}$, which shows two separate regions of magnetic hysteresis \cite{Yagasaki, Kadowaki, Nakama, Moskalenko, HuxleySANS} and a rich $M-T$ phase diagram with multiple magnetic field-induced transitions \cite{Nakama}. Furthermore, NMR/NQR, field-angle-resolved specific heat, and photoemission spectroscopy measurements \cite{Ishida, Manago, Kittaka, Kiss} all suggest an anisotropic $s$-wave superconducting gap symmetry and show that quasiparticle excitations are gapped-out at a finite temperature. The importance of 4$f$-electrons in this material has been highlighted by de~Haas~-~van~Alphen (dHvA) measurements \cite{Inada}, finding no change in frequency with the onset of superconductivity, but a cyclotron mass dependence consistent only with $f$-electron superconductors. Previous muon spin relaxation ($\mu$SR) research has proposed the presence of weak magnetism below $T_{M}$~=~40~K, a temperature much higher than the superconducting transition temperature ($T_c$~=~6~K) \cite{Huxley}. Polarized neutron experiments \cite{HuxleyPolN} reported field-induced paramagnetic moments on the order of 4.4$\times$10$^{-4}$~$\mu_B$T$^{-1}$ per Ce atom and 4.5$\times$10$^{-4}$~$\mu_B$T$^{-1}$ per Ru atom. The $\mu$SR technique was also used to probe the magnetic penetration depth within the high-field anomalous "peak effect" regime, suggesting the possibility of a Fulde-Ferrell-Larkin-Ovchinnikov (FFLO) state \cite{Yamashita,Kadono,KadonoRev}. However, no microscopic study of the superconducting gap symmetry and its interplay with magnetism is reported so far. 

\begin{figure*}[t!]
\centering
\includegraphics[width=1.1\linewidth]{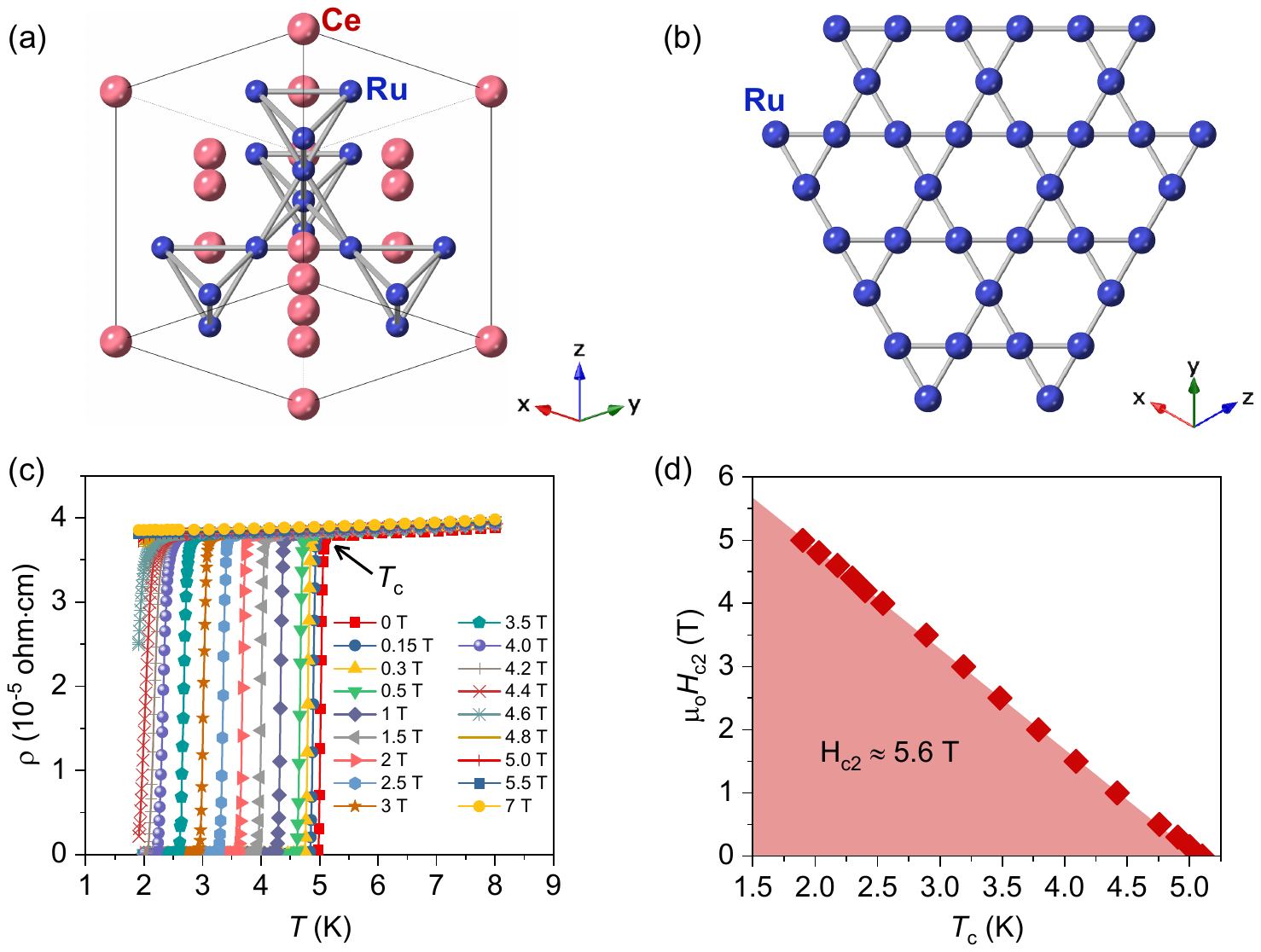}
\vspace{-0.85cm}
\caption{ (Color online)
(a) A three dimensional visualization of the atomic structure of CeRu$_{2}$. (b) When viewed along the [1~1~1] direction, a plane of Ru atoms constructs a pristine kagome lattice. (c) Resistivity measurements performed on a single crystal of CeRu$_2$ in different applied magnetic fields. The black arrow indicates the critical temperature, which was extracted at each field and used to construct panel (d), which shows the field dependence of the superconducting transition temperature. The data has been extrapolated with a straight line, giving a value of $H_{c2}$~$\simeq$~5.6~T at 1.5~K. }
\label{fig2}
\end{figure*}

 To provide a bulk local spectroscopic probe of the superconducting gap symmetry and of its interplay with weak magnetism in CeRu$_{2}$, we have carried out a combination of zero-field (ZF)-${\mu}$SR and high-field ${\mu}$SR experiments. Our studies were performed on two Czochralski-pulled single crystals, and the details of the sample preparation, additional characterization methods, and analysis techniques can be found in the Supplemental Information \cite{SuppInfo}. The superconducting order parameter achieves best agreement with an anisotropic $s$-wave symmetry. In the superconducting state, the ratio $T_{\rm c}$/$\lambda^{-2}$ (where $T_{\rm c}$ is the superconducting transition temperature and $\lambda^{-2}$ is the superfluid density) is comparable to those of unconventional superconductors. The relatively high $T_{\rm c}$ for the low carrier density may hint at an unconventional pairing mechanism in CeRu$_{2}$. The measured SC gap value $\Delta_{max}$~=~0.76(5)~meV yields a ratio 2$\Delta/k_{\rm B}T_{\rm c}$~${\simeq}$~3.8, suggesting that the superconductor CeRu$_{2}$ is in the moderate coupling limit. Furthermore, we identified three magnetic anomalies at $T_1^*$~${\simeq}$~110~K, $T_2^*$~${\simeq}$~65~K, and $T_3^*$~${\simeq}$~40~K. Importantly, these magnetic anomalies are strongly enhanced under a magnetic field of 8~T.

\section{Results and Discussion}

For the case of a material known to exhibit weak magnetism, $\mu$SR studies provide the most powerful tool for investigation. During a $\mu$SR experiment, positive muons are implanted into the sample, where they thermalize at interstitial positions and precess in the local magnetic field. They decay radioactively after a mean lifetime of 2.2~$\mu$s and emit a positron $e^+$ preferentially along the spin direction \cite{ZurabRev}. The asymmetry of emitted positrons is detected and this time-dependent polarization $P(t)$ of the ensemble may be fitted with a number of different functions, elucidating the physics of the investigated material (see the Supplemental Material \cite{SuppInfo} for details). The ${\mu}$SR technique provides a powerful tool to measure the magnetic penetration depth ${\lambda}$ in the vortex state (in the presence of a weak applied field $H_{\rm c1}~\leq H_{app}~\leq~H_{c2}$) of Type~II superconductors in the bulk of the sample \cite{GuguchiaMoTe2}, in contrast to many techniques that probe  ${\lambda}$ only near the surface. Additionally, zero-field ${\mu}$SR has the ability to detect internal magnetic fields as small as 0.1~G without applying external magnetic fields, making it a highly valuable tool for probing spontaneous magnetic fields due to time-reversal symmetry breaking.

Shown in Fig. 2a are the TF-$\mu$SR time spectra recorded in the normal state (10~K) and in the superconducting vortex state (0.27~K), measured on a single crystal of CeRu$_{2}$ with a field of 30~mT applied perpendicular to the $a$-axis. The cylindrical Czochralski-pulled sample, with $\varnothing~\simeq~6$ mm and a length of $\simeq~8$~ mm, was placed directly on the sample fork in the muon beam. Any muons not stopped in the sample passed through the aluminated mylar tape and were stopped in the veto detectors behind the sample; in this way, the additional signal of muons not stopped in the sample is immediately removed from the spectrum. The spectrum in the normal state shows a weak depolarization due to random local fields from the nuclear moments, which can be fitted by a single Gaussian distribution, while in the superconducting state the relaxation rate is strongly enhanced due to the formation of the flux-line lattice. As one can see in the field distribution (Fig.~2b), the profile is asymmetric and we fitted it with a sum of two Gaussian distributions. A single central field can then be extracted from the two Gaussian distributions, as detailed in the Supplementary Information \cite{SuppInfo}. The difference between the applied field (clearly visible as the center of the Gaussian field distribution in the sample in the normal state, see Fig.~2b at 10~K) and the central field in the superconducting state (see Fig.~2b at 0.27~K) constitutes the diamagnetic shift, plotted in Fig.~2d. Furthermore, the second moment of the two Gaussian field distributions can also be extracted and combined to obtain the relaxation rate \cite{SuppInfo} as displayed in Fig.~2c. From the temperature dependence of the diamagnetic shift ${\Delta}$$B_{\rm dia}$~=~${\mu}_{0}$($H_{\rm int,SC}$~-~$H_{\rm int,NS}$) (i.e., the difference between the internal field ${\mu}_{0}$$H_{\rm int,SC}$ measured in the SC fraction and ${\mu}_{0}$$H_{\rm int,NS}$ measured in the normal state at $T$ = 10 K), we can clearly see the large diamagnetic response of 0.7~mT associated with the superconducting transition at $T_c$~=~4.7~K in Fig.~2d. The temperature dependence of the superconducting muon spin depolarization rate, $\sigma_{\rm sc}$, is shown in Fig.~2c. In order to investigate the symmetry of the SC gap, we note that ${\lambda}(T)$ is related to the depolarization rate ${\sigma}_{{\rm sc}}(T)$ in the presence of a perfect triangular vortex lattice with $H_{\rm app} \ll H_{\rm c2}$ by the equation \cite{Brandt}: 
\begin{equation}
\frac{\sigma_{sc}(T)}{\gamma_{\mu}}=0.06091\frac{\Phi_{0}}{\lambda^{2}(T)},
\end{equation}
where ${\gamma_{\mu}}$ is the gyromagnetic ratio of the muon and ${\Phi}_{{\rm 0}}$ is the magnetic-flux quantum. 
The temperature dependence of the superfluid density $\lambda^{-2}(T)$ was then fitted with nodeless ($s$- and anisotropic $s$-wave) and nodal ($d$-wave) models to determine the superconducting gap symmetry \cite{SuppInfo}. 
Considering the quality of fit and $\chi_r^2$ values (see Fig. 2c and Table 1), it is clear that the anisotropic $s$-wave fits the data best, meaning that there is an angular dependence to the superconducting gap value (similar to $d$-wave superconductivity) but the minimum gap value is nonzero. The ratio of the minimum gap value to the maximum gap value we obtained was $\Delta_{min}/\Delta_{max}$~=~0.47(1). This is in fairly good agreement with values near $\Delta_{min}$/$\Delta_{max}$~=~0.33 and $\Delta_{min}$/$\Delta_{max}$~=~0.20 obtained from NMR studies \cite{Kittaka} and in excellent agreement with the value obtained by photoemission experiments $\Delta_{min}$/$\Delta_{max}$~=~0.447 \cite{Kiss}. 

\begin{figure*}[t!]
\centering
\includegraphics[width=1.0\linewidth]{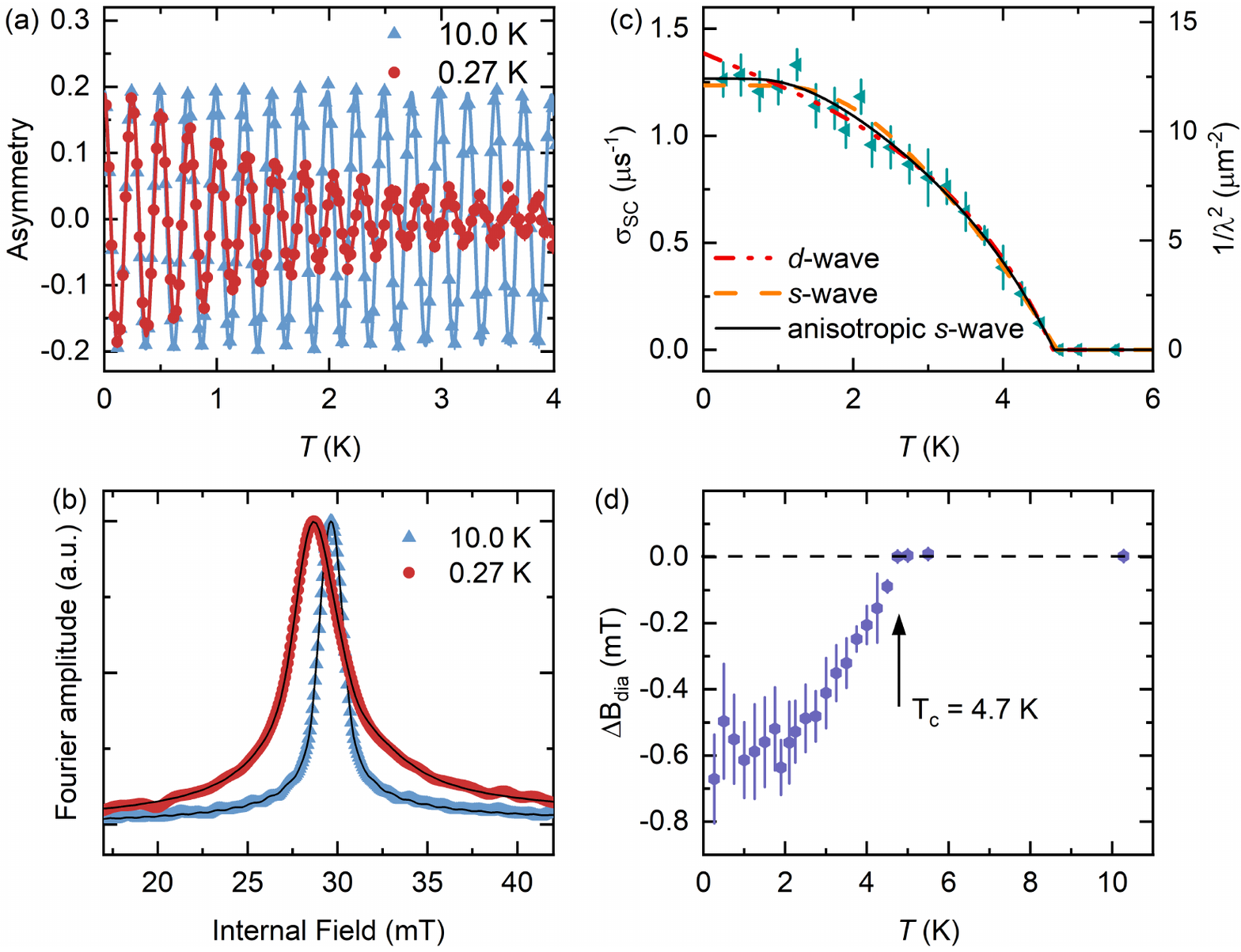}
\vspace{0cm}
\caption{ Color online)
(a) Transverse-field (TF) ${\mu}$SR time spectra of CeRu$_{2}$ probing the superconducting vortex state. The TF spectra are obtained above and below $T_{\rm c}$ (after field cooling the sample from above $T_{\rm c}$). Details of the fitting procedures can be found in \cite{SuppInfo}. (b) Fourier transforms of the ${\mu}$SR time spectra shown in panel (a). Temperature dependence of the muon spin depolarization rate ${\sigma}_{\rm sc}$($T$) (c) and the diamagnetic shift ${\Delta}$$B_{\rm dia}$, measured in an applied magnetic field of ${\mu}_{\rm 0}H~=~30$~mT. (d) The arrow marks the $T_{c}$ value seen clearly from the diamagnetic shift.}
\label{fig2}
\end{figure*}

\begin{table}
\caption{Summary of the parameters obtained for fits of the superconducting gap structure to the superfluid density measured in CeRu$_{2}$ by means of ${\mu}$SR. The reduced $\chi^2$ values indicated in the rightmost column clearly show the best fit is obtained by the anisotropic $s$-wave gap structure. The fitting procedure and functions used can be found in \cite{SuppInfo}. }
\begin{center}
\begin{tabular}{| c | c | c | c | c | c |}
\hline 
Symmetry & $\lambda_0$ [nm] & $T_{c}$ [K] & $\Delta_{max}$ [meV] & $\Delta_{min}$ [meV] & $\chi_r^{2}$\\
\hline
an. $s$-wave & 284(5) & 4.68(3) & 0.76(5) & 0.36(1) & 0.76\\
$s$-wave & 287(4) & 4.72(3) & 0.73(4) & - & 1.05\\
$d$-wave & 271(4) & 4.67(2) & 1.12(7) & - & 0.85\\
\hline 
\end{tabular}
\end{center}
\end{table}

\begin{figure*}[t!]
\centering
\includegraphics[width=1.0\linewidth]{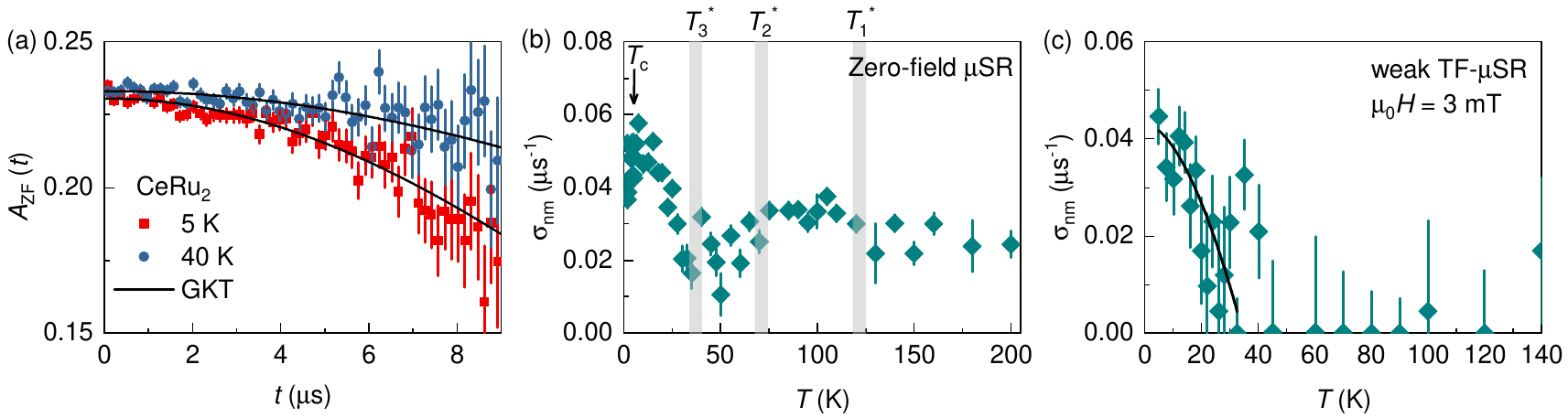}
\vspace{-0.7cm}
\caption{ (Color online)  
(a) ZF ${\mu}$SR time spectra for CeRu$_{2}$ recorded at 5~K and 40~K. The line represents the fit to the data using a standard Kubo-Toyabe depolarization function \cite{Toyabe}, reflecting the field distribution at the muon site created by the nuclear moments. Temperature dependence of the muon spin Gaussian depolarization rate ${\sigma}_{\rm nm}$, measured in precise zero-field (b) and small magnetic field of 3~mT, applied perpendicular to the muon spin polarization (c).}
\label{fig4}
\end{figure*}

  From the measured muon relaxation rate in the superconducting state, we can calculate the superfluid density using Equation~1. The ratio of the superconducting gap to $T_{\rm c}$ was estimated to be 2$\Delta_{max}/k_{\rm B}T_{\rm c}$~=~3.8, which is in excellent agreement with NMR, dHvA effect, photoemission, surface impedance, and tunneling results  \cite{Ishida, Inada, Kiss, Matsui, Moskalenko}. This ratio is consistent with the moderate coupling limit BCS expectation \cite{GuguchiaMoTe2}. However, a similar ratio can also be expected for the Bose Einstein Condensate (BEC)-like picture as pointed out in ref. \cite{Uemura4}. The Uemura ratio \cite{Uemura1} between the critical temperature and the superfluid density extrapolated to $T$~=~0~K is estimated to be $T_c$/$\lambda^{-2}$~$\simeq$~0.377, which is an order of magnitude smaller than for hole-doped cuprate superconductors, but still far away from conventional phonon-mediated BCS superconductors \cite{GuguchiaNbSe2}. Interestingly, the ratio for CeRu$_{2}$ is almost identical to that for LaRu$_3$Si$_2$ \cite{MielkeLRS}, for charge density wave superconductors 2H-NbSe$_{2}$ and 4H-NbSe$_{2}$ as well as Weyl-superconductor $T_{d}$-MoTe$_{2}$ \cite{GuguchiaMoTe2}. This finding hints at an unconventional pairing mechanism in CeRu$_{2}$ with a low density of Cooper pairs and similar electron correlations as in LaRu$_3$Si$_2$, 2H-NbSe$_{2}$ and $T_{d}$-MoTe$_{2}$, but much weaker electron correlations than in cuprates and twisted bilayer graphene. 

While the unconventional nature itself of this superconductor makes it an interesting subject, a previous $\mu$SR study \cite{Huxley} found extremely weak magnetism in this material at a temperature much higher than the superconducting transition, $T_{M}$~=~40~K. Motivated by the apparent similarity to recently-discovered KV$_{3}$Sb$_{5}$, in which the onset of a charge density wave phase is accompanied by electron dynamics that break time-reversal symmetry \cite{MielkeKVS}, we similarly performed ZF-$\mu$SR measurements over a broad temperature range on CeRu$_{2}$. We observe a clear increase in the relaxation rate, evidenced by the comparison of the $\mu$SR spectra observed at 5~K and 40~K, in Fig.~3a. The $\mu$SR spectra were fitted with a Gaussian Kubo-Toyabe function. It has been previously shown that the muon spin relaxation originates from a static internal field distribution, as a longitudinal field of 1~mT is sufficient to fully decouple the depolarization \cite{Huxley}. We notice an upturn and a broad downturn with the onsets of $T_1^*$~$\sim$~110~K and $T_2^*$~$\sim$~65~K. Consistent with the earlier report \cite{Huxley}, we also notice a small increase of 0.03~$\mu$s$^{-1}$ in ${\sigma}_{nm}$ around 40~K, which we have denoted as $T_3^*$ in Fig.~3b. With the application of 3~mT, we can more clearly identify the enhancement below 40~K, as seen in Fig.~3c. It is interesting to note the reduction of zero-field rate ${\sigma}_{nm}$ below the superconducting transition temperature $T_{\rm c}$ (see Fig.~3b). This indicates a clear effect of superconductivity on the weak internal fields and supports the magnetic origin of the increased depolarization rate. More importantly, this behavior indicates an interplay between magnetism and superconductivity in CeRu$_{2}$ involving competition for the same electrons. The strong suppression of the magnetism below the onset of superconductivity was also observed in the nodeless Fe-based high-temperature superconductors: NaFe$_{1-x}$Ni$_{x}$As \cite{Cheung}, BaFe$_{2-x}$Co$_{x}$As$_{2}$ \cite{Goltz,Tam},  BaFe$_{2-x}$Ni$_{x}$As$_{2}$ \cite{Arguello}, Ba$_{1-x}$K$_{x}$Fe$_{2}$As$_{2}$ \cite{Wiesenmayer}, and FeSe \cite{Bendele}. It was discussed that itinerant AFM and SC orders are generally expected to compete strongly for the same electronic states, which was captured within a simple Ginzburg-Landau free energy for the AFM and SC order parameters \cite{Cheung,Fernandes}.
 
\begin{figure}[t!]
\centering
\includegraphics[width=1.0\linewidth]{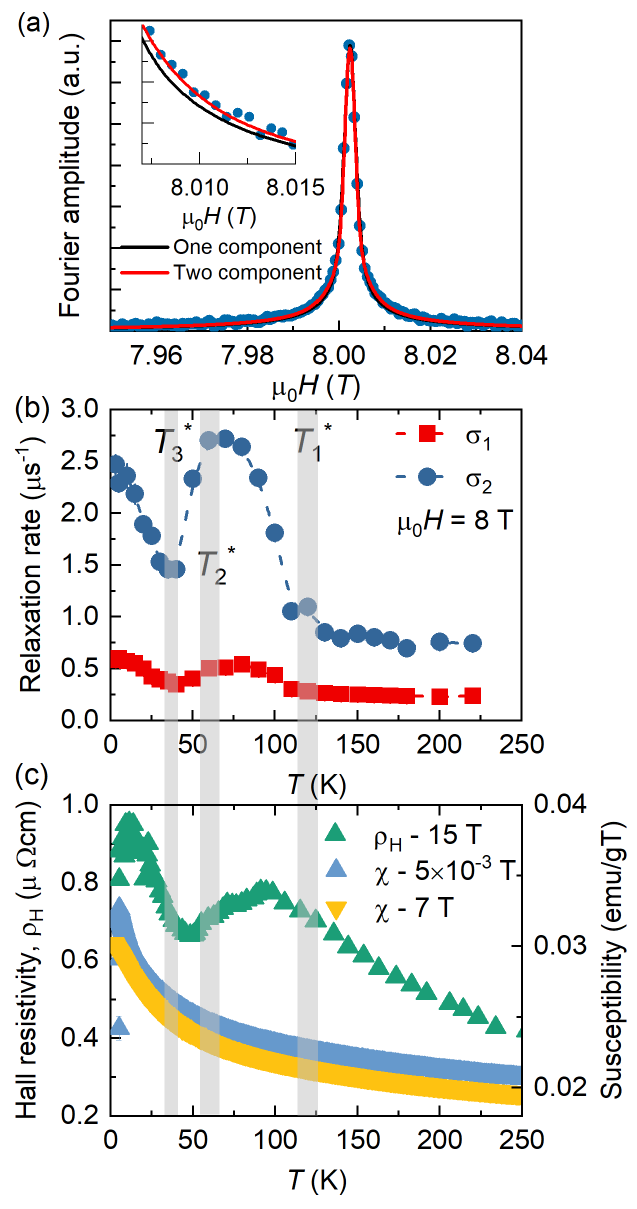}
\vspace{-0.3cm}
\caption{ (Color online)  
(a) Fourier transform for the ${\mu}$SR asymmetry spectra of CeRu$_{2}$ at 5~K for the applied field of ${\mu}_{0}$$H$~=~8~T. The black and red solid lines represent the fits to the data using the one and two component signals, respectively. (b) The temperature dependences of the Gaussian muon spin relaxation rates ${\sigma}_{\rm 1}$ and ${\sigma}_{\rm 2}$. (c)  (Left axis) The temperature dependence of the Hall resistivity, measured in 15~T (after ref. \cite{Nakama}). (Right axis) The temperature dependence of the magnetic susceptibility, measured in 0.005~T and 7~T.}
\label{fig4}
\end{figure}

In order to confirm the magnetic origin of the low-$T$ relaxation rate, we performed high-field $\mu$SR experiments with the HAL-9500 instrument in 8~T applied along the cylindrical cut of the sample. The magnetic contribution should experience field-induced enhancement, but the nuclear contribution should remain unchanged \cite{MielkeKVS}. Since the upper critical field in this superconductor is $H_{c2}$~$\approx$~5.6~T \cite{Inada, Yagasaki, HuxleySANS}, we were able to perform the $\mu$SR experiments purely in the normal state down to the base temperature of 3~K with complete suppression of the superconducting state. As illustrated in Fig.~4a, the high-field $\mu$SR spectra are best described by a two-component Gaussian fit. We observed that a single-component fit was not sufficient to describe the field distribution in CeRu$_2$ under applied field, and that a second Gaussian component was needed to fit the $\mu$SR spectra, as illustrated in Fig.~4a. The contribution seen in Fig.~4b from the $\sigma_2$ component (ie. the blue circles) accounts only for 30\% of the total signal. These two components and their relative fractions may be related to the presence of both Ce$^{+3}$ and Ce$^{+4}$ states as evidenced by photoemission experiments \cite{Sekiyama} and crystallographic muon sites with different relative distances to the corresponding ions. Another possibility for the presence of two components is a phase separation into two spatially separated volumes with different magnetic properties. In Fig.~4b is displayed the temperature dependence of both components, which both show clear anomalies at all three critical temperatures identified. The dome-like feature, which begins to increase around $T_1^*$~$\sim$~110~K (concomitant with the onset of the increase observed with ZF-$\mu$SR in Fig.~3b), reaches a maximum at $T_2^*$~$\sim$~65~K, below which the rate decreases. Both relaxation rates begin sharply increasing again at $T_3^*$~=~40~K, coincident with the increase observed by ZF-$\mu$SR (see Fig.~3a). Most importantly, the observed anomalies are strongly enhanced under applied magnetic field. Namely, the relaxation rate $\sigma_2$ increases by 1.5 and 1~$\mu$s$^{-1}$ below $T_1^*$ and $T_3^*$, respectively, in 8~T. This is two orders of magnitude higher than the increase of 0.03~$\mu$s$^{-1}$ observed in zero-field. This clearly supports the electronic/magnetic origin of the anomalies seen under zero-field, as the temperature dependence of the nuclear contribution to the relaxation cannot be significantly changed by an external field. It is also noteworthy to mention that the previously reported high-field Hall resistivity (being highly sensitive to magnetic contributions) exhibits very similar temperature dependence as our high-field ${\mu}$SR relaxation rate (see Fig.~4c). 

The combination of ZF-${\mu}$SR and high-field ${\mu}$SR results on CeRu$_{2}$ provides evidence of distinct magnetic responses with three characteristic temperatures $T_{{\rm 1}}^{*}$~${\simeq}$~110~K, $T_{{\rm 2}}^{*}$~${\simeq}$~65~K, and $T_{{\rm 3}}^{*}$~${\simeq}$~40~K. This may originate from the complex interaction between Ru-$d$-electrons and Ce-4$f$ electrons \cite{Sekiyama,JSKang}. We can not comment on the precise origin of magnetism in CeRu$_{2}$. However, since macroscopic susceptibility does not show any clear magnetic transitions (see Fig.~4c), magnetism is likely itinerant and antiferromagnetic. This calls for additional detailed experiments.

 The presence of weak magnetism in CeRu$_{2}$ is reminiscent of kagome superconductors KV$_{3}$Sb$_{5}$ and RbV$_{3}$Sb$_{5}$ \cite{MielkeKVS,GuguchiaRVS}, where ${\mu}$SR shows the emergence of a time-reversal symmetry-breaking state below 75~K and 120~K, respectively. However, in the 135 kagome superconductors, the weak magnetic signal occurs contemporaneously with topological charge ordering, which competes with superconductivity \cite{GuguchiaRVS}, ocurring at much lower temperatures $T_{{\rm c}}$~${\simeq}$~1~K. The $T_{{\rm c}}$ of (K,Rb)V$_{3}$Sb$_{5}$ are enhanced to ${\simeq}$~4~K under pressure, only after suppressing the charge order. Furthermore, the superconducting pairing symmetry is nodal for both (K,Rb)V$_{3}$Sb$_{5}$ at low pressure when the system also exhibits charge order \cite{GuguchiaRVS}. Upon applying pressure, the charge order is suppressed and the superconducting state progressively evolves from nodal to nodeless \cite{GuguchiaRVS}. Thus, the high-pressure SC state in (K,Rb)V$_{3}$Sb$_{5}$ without charge order is nodeless. No charge ordering has been reported for CeRu$_{2}$ even at ambient pressure and it exhibits a nodeless superconducting state with a relatively high critical temperature $T_{{\rm c}}$~${\simeq}$~5~K, similar to kagome superconductor LaRu$_{3}$Si$_{2}$. All these observations strongly suggest that the presence of charge order in kagome superconductors can strongly influence the superconducting gap structure.\\ 
 
\section{Conclusion}

The distorted Laves-phase $f$-electron superconductor CeRu$_{2}$ exhibits a pristine Ru kagome network, which has been shown to host correlated electronic states. Using the bulk-sensitive magnetic microprobe $\mu$SR, we have spectroscopically identified CeRu$_2$ as a nodeless superconductor, with a temperature dependence of the superconducting order parameter which is best fitted by an anisotropic $s$-wave gap symmetry. The unconventional nature of superconductivity is additionally evidenced by the observed dilute superfluid density. Furthermore, the combination of highly-sensitive ZF-$\mu$SR and high-field $\mu$SR shows that this material exhibits a magnetic response with three characteristic temperatures, which we have identified as $T_1^*$~=~110~K, $T_2^*$~=~65~K and $T_3^*$~=~40~K. We furthermore show that the magnetic response is strongly enhanced by magnetic field. Our bulk spectroscopic characterization of the nodeless kagome superconductivity and magnetic order underline the competition between these two orders in CeRu$_2$. \\

\section{Acknowledgments}~
The ${\mu}$SR experiments were carried out at the Swiss Muon Source (S${\mu}$S) Paul Scherrer Insitute, Villigen, Switzerland. 
Z.G. acknowledges useful discussions with Dr. Robert Johann Scheuermann. Y.S. acknowledges the support from the National Natural Science Foundation of China (Grants No. U2032204),  and the K. C. Wong Education Foundation (GJTD-2018-01). M.Z.H. acknowledges visiting scientist support from IQIM at the California Institute of Technology. 
C.W. Chu acknowledges support from US Air Force Office of Scientific Research Grants FA9550-15-1-0236 and FA9550-20-1-0068, the T.L.L. Temple Foundation, the John J. and Rebecca Moores Endowment, and the State of Texas through the Texas Center for Superconductivity at the University of Houston. The magnetization measurements were carried out on the MPMS device of the Laboratory for Multiscale Materials Experiments, Paul Scherrer Institute, Villigen, Switzerland (SNSF grant No. 206021-139082) and on the MPMS3 device at the UZH in Z\"{u}rich, Switzerland (SNSF grant No. 206021-150784).


\begin{thebibliography}{150}

\bibitem{GuguchiaCSS} Z. Guguchia et al.
Tunable anomalous Hall conductivity through volume-wise magnetic competition in a topological kagome magnet.
Nature Communications \textbf{11}, 559 (2020).

\bibitem{JXYin2} J-X. Yin et. al. 
Giant and anisotropic spin-orbit tunability in a strongly correlated kagome magnet. 
Nature \textbf{562}, 91-95 (2018).

\bibitem{MielkeKVS} C. Mielke III et al.
Time-reversal symmetry-breaking charge order in a kagome superconductor.
Nature \textbf{602}, 245-250 (2022).

\bibitem{JiangpingHu} K. Jiang, T. Wu, J.-X. Yin, Z. Wang, M.Z. Hasan, S.D. Wilson, X. Chen, and J. Hu.
Kagome superconductors $A$V$_{3}$Sb$_{5}$ ($A$=K, Rb, Cs).
Preprint at https://arXiv:2109.10809 (2021).

\bibitem{TNeupert} T. Neupert, M.M. Denner, J.-X. Yin, R. Thomale, and M.Z. Hasan.
Charge order and superconductivity in kagome materials.
Nature Physics {\bf 18}, 137-143 (2022).

\bibitem{Yu} S.-L. Yu and J.-X. Li.
Chiral superconducting phase and chiral spin-density-wave phase in a Hubbard model on the kagome lattice.
Phys. Rev. B \textbf{85}, 144402 (2012).

\bibitem{Ortiz} B. Ortiz et al.
CsV$_3$Sb$_5$: A Z$_2$ topological kagome metal with a superconducting ground state.
Phys. Rev. Lett. \textbf{125}, 247002 (2020).

\bibitem{MielkeLRS} C. Mielke III et al.
Nodeless kagome superconductivity in LaRu$_3$Si$_2$.
Phys. Rev. Materials \textbf{5}, 034803 (2021).

\bibitem{BOrtiz3} B. Ortiz, P. Sarte, E. Kenney, M. Graf, S. Teicher, R. Seshadri, \& S. Wilson.
Superconductivity in the Z$_{2}$ kagome metal KV$_{3}$Sb$_{5}$.
\textit{Phys. Rev. Materials} {\bf 5}, 034801 (2021).

\bibitem{YJiang} Y.-X. Jiang et al. 
Discovery of topological charge order in kagome superconductor KV$_{3}$Sb$_{5}$.
\textit{Nature Materials} {\bf 20}, 1353-1357 (2021).

\bibitem{GuguchiaRVS} Z. Guguchia et al.
Tunable nodal kagome superconductivity in charge ordered RbV$_{3}$Sb$_{5}$.
Preprint at https://arXiv:2202.07713 (2022).

\bibitem{NShumiya} N. Shumiya et al.
Tunable chiral charge order in kagome superconductor RbV$_{3}$Sb$_{5}$.
\textit{Phys. Rev. B} {\bf 104}, 035131 (2021).

\bibitem{Ritu} R. Gupta et al. 
Microscopic evidence for anisotropic multigap superconductivity in the CsV$_3$Sb$_5$ kagome superconductor.
npj Quantum Materials \textbf{7}, 49 (2022). 

\bibitem{HuxleyPolN} A. Huxley et al.
The magnetic and crystalline structure of the Laves phase superconductor CeRu$_2$.
J. Phys.: Cond. Matter \textbf{9}, 4185-4195 (1997).

\bibitem{Deng} L.Z. Deng et al.
Magnetic kagome superconductor CeRu$_{2}$.
Preprint at https://arxiv.org:2204.00553 (2022).

\bibitem{Sekiyama} A. Sekiyama, T. Iwasaki, K. Matsuda, Y. Saitoh, Y. Onuki, S. Suga.
Probing bulk states of correlated electron systems by high-resolution resonance photoemission.
Nature \textbf{403}, 396-398 (2000).

\bibitem{JSKang} J.-S. Kang, C.G. Olson, M. Hedo, Y. Inada, E. Yamamoto, Y. Haga, Y. Onuki, S.K. Kwon, B.I. Min. 
Photoemission study of an $f$-electron superconductor: CeRu$_{2}$.
Phys. Rev. B \textbf{60}, 5348-5353 (1999).

\bibitem{Kadowaki} K. Kadowaki, H. Takeya, K. Hirata. 
Anomalous magnetization behavior of single-crystalline CeRu$_2$.
Phys. Rev. B \textbf{54}, 462 (1996).

\bibitem{Yagasaki} K. Yagasaki, M. Hedo, T. Nakama. 
Reentrant superconductivity of CeRu$_2$.
J. Phys. Soc. Japan \textbf{62}, 3825-3828 (1993).

\bibitem{Nakama} T. Nakama, M. Hedo, T. Maekawa, M. Higa, R. Resel, H. Sugawara, R. Settai, Y. Onuki, K. Yagasaki.
Evidence of multiple superconducting phases in CeRu$_2$.
J. Phys. Soc. Japan \textbf{64}, 1471-1475 (1995).

\bibitem{HuxleySANS} A. Huxley et al.
A neutron study of the flux lattice in the superconductor CeRu$_2$.
Physica B \textbf{223\&224}, 169-171 (1996).

\bibitem{Moskalenko}  A. Moskalenko, Y. Naidyuk, I. Yanson, M. Hedo, Y. Inada, Y. Onuki, Y. Haga, E. Yamamoto. 
Superconducting gap and pair breaking in CeRu$_2$ studied by point contacts. 
Fiz. Nizk. Temp. \textbf{27}, 831-834 (2001).

\bibitem{Ishida} K. Ishida, H. Mukuda, Y. Kitaoka, K. Asayama, Y. Onuki. 
Ru NMR and NQR Studies in CeRu$_2$.
Z. Naturforsch. \textbf{51a}, 793-796 (1996). 

\bibitem{Kiss} T. Kiss et al.
Photoemission spectroscopic evidence of gap anisotropy in an $f$-electron superconductor. 
Phys. Rev. Lett. \textbf{94}, 057001 (2005). 

\bibitem{Kittaka} S. Kittaka, T. Sakakibara, M. Hedo, Y. Onuki, K. Machida.
Verification of anisotropic $s$-wave superconducting gap structure in CeRu$_2$ from low-temperature field-angle-resolved specific heat measurements.
J. Phys. Soc. Japan \textbf{82}, 123706 (2013).

\bibitem{Manago} M. Manago, K. Ishida, T. Matsuda, Y. Onuki.
Effect of geomagnetism on $^{101}$Ru Nuclear Quadrupole Resonance measurements of CeRu$_2$.
J. Phys. Soc. Japan \textbf{84}, 115001 (2015).

\bibitem{Inada} Y. Inada and Y. Onuki. 
De Haas - van Alphen oscillation in both the normal and superconducting mixed states of NbSe$_2$, CeRu$_2$, URu$_2$Si$_2$, and UPd$_2$Al$_3$.
Low Temp. Physics \textbf{25}, 573 (1999).

\bibitem{Huxley} A. Huxley, P. Dalmas de Reotier, A. Yaouanc, D. Caplan, M. Couach, P. Lejay, P. Gubbens, A. Mulders. 
CeRu$_{2}$: A magnetic superconductor with extremely small magnetic moments. 
Phys. Rev. B \textbf{54}, R9666(R) (1996).

\bibitem{Yamashita} A. Yamashita et al. 
Anomalous field dependence of magnetic penetration depth in the vortex state of CeRu$_2$ probed by muon spin rotation.
Phys. Rev. Letters \textbf{79}, 3771 (1997).

\bibitem{Kadono} R. Kadono et al. 
Possible nodal vortex state in CeRu$_2$.
Phys. Rev. B \textbf{63}, 224520 (2001).

\bibitem{KadonoRev} R. Kadono.
Field-induced quasiparticle excitations in novel type II superconductors.
J. Phys.: Condens. Matter \textbf{16}, S4421 (2004).

\bibitem{SuppInfo} See Supplemental Information for the details of sample preparation, additional experiments, and $\mu$SR experiments and data analysis procedures.

\bibitem{ZurabRev} Z. Guguchia.
Unconventional magnetism in layered transition metal dichalcogenides.
Condens. Matter \textbf{5}, 42 (2020).


\bibitem{GuguchiaMoTe2} Z. Guguchia et al. 
Signatures of the topological $s^{+-}$ superconducting order parameter in the type-II Weyl semimetal $T_{d}$-MoTe$_{2}$.
Nature Communications \textbf{8}, 1082 (2017).


\bibitem{Brandt} E.H. Brandt. 
Flux distribution and penetration depth measured by muon spin rotation in high-$T_{{\rm c}}$ superconductors.
$Phys.~Rev.~B$ \textbf{37}, 2349 (1988).


\bibitem{Matsui} H. Matsui, T. Yasuda, M. Hedo, R. Settai, Y. Onuki, E. Yamamoto, Y. Haga, N. Toyota. 
Electrodynamics of the Superconducting State in CeRu$_2$.
J. Phys. Soc. Jpn. \textbf{67}, 3580 (1998).


\bibitem{Uemura4} Y.J. Uemura et. al.
Basic Similarities among Cuprate, Bismuthate, Organic, Chevrel Phase, and Heavy-Fermion Superconductors Shown by Penetration Depth Measurements.
Phys. Rev. Lett. \textbf{66}, 2665 (1991).

\bibitem{Uemura1} Y.J. Uemura et al. 
Universal correlations between $T_c$ and $\frac{n_s}{m^*}$ (carrier density over effective mass) in high-$T_c$ cuprate superconductors.
Phys. Rev. Lett. \textbf{62}, 2317 (1989).

\bibitem{GuguchiaNbSe2} F. O. von Rohr et al. 
Unconventional Scaling of the Superfluid Density with the Critical Temperature in Transition Metal Dichalcogenides.
Science Advances 5(11), eaav8465 (2019).

\bibitem{Cheung} S.C. Cheung et al.
Disentangling superconducting and magnetic orders in NaFe$_{1?x}$Ni$_{x}$As using muon spin rotation.
Phys. Rev. B \textbf{97}, 224508 (2018).

\bibitem{Goltz} T. Goltz, V. Zinth, D. Johrendt, H. Rosner, G. Pascua, H. Luetkens, P. Materne, and H.-H. Klauss. 
Microscopic coexistence of magnetism and superconductivity in charge-compensated Ba$_{1-x}$K$_x$(Fe$_{1-y}$Co$_y$)$_2$As$_2$.
Phys. Rev. B \textbf{89}, 144511 (2014).

\bibitem{Tam} D.W. Tam et al.
Uniaxial pressure effect on the magnetic ordered moment and transition temperatures in BaFe$_{2-x}T_x$As$_2$ ($T$ = Co, Ni).
Phys. Rev. B \textbf{95}, 060505 (2017).

\bibitem{Arguello} C.J. Arguello. 
Ph.D. thesis, Columbia University, 2014.

\bibitem{Wiesenmayer} E. Wiesenmayer, H. Luetkens, G. Pascua, R. Khasanov, A. Amato, H. Potts, B. Banusch, H.-H. Klauss, and D. Johrendt.
Microscopic coexistence of superconductivity and magnetism in Ba$_{1-x}$K$_x$Fe$_2$As$_2$.
Phys. Rev. Lett. \textbf{107}, 237001 (2011).

\bibitem{Bendele} M. Bendele et al.
Coexistence of superconductivity and magnetism in FeSe$_{1-x}$ under pressure.
Phys. Rev. B \textbf{85}, 064517 (2012).

\bibitem{Fernandes} R.M. Fernandes and J. Schmalian. 
Competing order and nature of the pairing state in the iron pnictides.
Phys. Rev. B \textbf{82}, 014521 (2010).

\bibitem{Bastian} A. Suter and B.M. Wojek. 
Musrfit: a free platform-independent framework for $\mu$SR data analysis.
Physics Procedia \textbf{30}, 69 (2012).

\bibitem{Toyabe} R. Kubo and T. Toyabe. 
Magnetic Resonance and Relaxation (North Holland, Amsterdam, 1967).

\bibitem{LNie} L. Nie, et. al. 
Charge-density-wave-driven electronic nematicity in a kagome superconductor.
Nature \textbf{604}, 59 (2022).


\end{thebibliography}
\end{document}


\preprint{preprint(\today)}

\title{Local spectroscopic evidence for a nodeless magnetic kagome superconductor CeRu$_{2}$}

\section{Supplementary Information}

\textbf{General remark}: Here, we concentrate on muon spin rotation/relaxation/resonance ($\mu$SR) \cite{Sonier,Brandt,GuguchiaNature} measurements of the magnetic penetration depth $\lambda$ in CeRu$_{2}$, which is one of the fundamental parameters of a superconductor, since it is related to the superfluid density $n_{s}$ via 1/${\lambda}^{2}$ = $\mu_{0}$$e^{2}$$n_{s}/m^{*}$ (where $m^{*}$ is the effective mass). Most importantly, the temperature dependence of ${\lambda}$ is particularly sensitive to the structure of the SC gap. Moreover, zero-field ${\mu}$SR is a very powerful tool for detecting static or dynamic magnetism in exotic superconductors, because very small internal magnetic fields are detected in measurements without applying external magnetic fields.\\ 

\textbf{Sample preparation}: A single crystal of CeRu$_{2}$ was grown using the Czochralski pulling method with a prepared polycrystalline ingot in a tetra-arc furnace. Firstly, a spheroidal ingot with a mass of 10~g was prepared by arc melting stoichiometric amounts of lumpish cerium and ruthenium, which was then placed in the water-cooled copper hearth of the tetra-arc furnace under argon atmosphere with a titanium ingot used as an oxygen getter. A current of 16~A was applied to the ingot to melt it and the velocity of the hearth rotation was 0.3~revolutions per minute. A columnar single crystal was successfully grown after several hours. As CeRu$_2$ crystallizes in the cubic $F\overline{4}3m$ space group \cite{HuxleyPolN}, it did not grow along one of the principal crystallographic axes; rather, by laboratory Laue diffraction, we determined that the cylindrical single crystal grew along the (0~2~1) direction. For the ZF-${\mu}$SR and superfluid density measurements, the entire cylindrical sample was mounted on the sample holder perpendicular to the beam; thus, the incident muon beam was perpendicular to the (0~2~1) crystallographic direction. For the high field ${\mu}$SR, a cut of the sample was mounted with the field along the (0~2~1) direction. The focus of this this study was not on the anisotropic properties of CeRu$_2$; however, this will be the subject of future studies.\\

\textbf{Electrical transport measurement:} Resistivity measurements were performed by using the Quantum Design Physical Property Measurement System (PPMS) under different magnetic fields up to 7 T and at temperatures down to 1.8 K.\\ 

\textbf{${\mu}$SR experiment}:  In a ${\mu}$SR experiment, nearly 100~${\%}$ spin-polarized muons ${\mu}$$^{+}$ are implanted into the sample one at a time. The incoming muons (with momentum $p_{\mu}$~=~29~MeV/c) are stopped in the sample, which can have an applied transverse field (TF) or zero field (ZF) condition. The positively charged ${\mu}$$^{+}$ thermalize at interstitial lattice sites, where they act as magnetic microprobes. In a magnetic material, the muon spin precesses in the local field $B_{\rm \mu}$ at the muon site with the Larmor frequency ${\omega}_{\rm \mu}$ = $\gamma_{\rm \mu}$/(2${\pi})$$B_{\rm \mu}$ (muon gyromagnetic ratio $\gamma_{\rm \mu}$/(2${\pi}$) = 135.5 MHz T$^{-1}$). In a ZF $\mu$SR experiment, the implanted positive muons precess in any magnetic field at the muon stopping site, allowing them to serve as an extremely sensitive local probe to detect small internal fields and ordered magnetic volume fractions in the bulk of magnetic materials.

The positive muons $\mu^+$ which are implanted in the sample decay after a mean life time of $\tau_{\mu}$~=~2.2~$\mu$s, and emit a fast position $e^+$ preferentially along their spin direction. This process is tracked by detectors surrounding the sample; an incoming muon detector starts an electronic clock when the $\mu^+$ passes, which is stopped by the detection of the associated decay position $e^+$. Such a decay event must occur in a certain time interval, usually over several muon lifetimes (e.g. 10 $\mu$s). The measured time interval is stored in a histogramming memory, forming a histogram of positron count versus time. The GPS (General Purpose Surface-Muon) instrument uses two detector pairs surrounding the sample; Forward, Backward, Left, and Right. After several millions of muons are stopped in the sample, one at a time, one obtains a histogram for the positrons $e^+$ collected on the Forward ($N_F$), Backward ($N_B$), Left ($N_L$), and Right ($N_R$) detectors. The histogram counts are ideally described by:
\begin{equation}
N_{\alpha}\left( t \right) = N_0e^{-\frac{t}{\tau_{\mu}}}\left[1+A_0\vec{P}(t) \hat{n}_{\alpha} \right] + N_{\rm bg}
\end{equation} 
where the exponential factor accounts for the radioactive muon decay; $\vec{P}\left(t\right)$ is the muon-spin polarization function with the unit vector $\vec{n}_{\alpha}$ ($\alpha$ = F, B, L, R) along the direction of the positron detector; $N_0$ is a parameter proportional to the number of recorded events; $N_{\rm bg}$ is the background contribution due to uncorrelated starts and stops; and $A_0$ is the initial decay asymmetry, which depends on a number of experimental factors such as the detector solid angle, absorption, and the scattering of positrons in the material. Typical values of $A_0$ are between 0.2 and 0.3 \cite{ZurabRev, GuguchiaStrain}. 

As the positrons are emitted predominantly in the direction of the muon spin which precesses with $\omega_\mu$, the Forward and Backward detectors detect a signal oscillating at the same frequency. The exponential decay (which simply reflects the muon's finite lifetime) can be removed by calculating the asymmetry signal, $A(t)$:
\begin{equation}
A\left( t \right) = \frac{N_{F,L}\left(t\right)-N_{B,R}\left(t\right)}{N_{F,L}\left(t\right)+N_{B,R}\left(t\right)} = A_0P\left(t\right)
\end{equation}
where $N_{F,L}\left(t\right)$ and $N_{B,R}\left(t\right)$ are the number of positrons detected by the Forward(Left) and Backward(Right) detectors, respectively. The quantities $A(t)$ and $P(t)$ depend sensitively on the static spatial distribution and the fluctuations of the magnetic environment of the muons. These functions reflect the physics of the investigated system \cite{GuguchiaStrain,ZurabRev}.

Using the $\mu$SR technique, important length scales of superconductors can be measured, namely the magnetic penetration depth $\lambda$ and the coherence length $\xi$. If a Type II superconductor is cooled below $T_{\rm c}$ in an applied magnetic field ranging between the lower ($H_{c1}$) and the upper ($H_{c2}$) critical fields, a vortex lattice is formed which in general is incommensurate with the crystal lattice, with vortex cores separated by much larger distances than those of the crystallographic unit cell. Because the implanted muons stop at given crystallographic sites, they will randomly probe the field distribution of the vortex lattice. Such measurements need to be performed in a field applied perpendicular to the initial muon spin polarization (so-called TF configuration). \\
 
\textbf{Fitting of the ZF-$\mu$SR data}:
The $\mu$SR spectra collected in zero field with active compensation (in which even geomagnetism is compensated) were fitted over the entire temperature range using the following function:
\begin{equation}
A_{\rm ZF}=A_0 \left( \frac{1}{3} + \frac{2}{3} \left[ 1 - (\sigma t)^2\right] e^{-\frac{1}{2} (\sigma t)^2} \right)
\end{equation}
which is also known as the Gaussian Kubo-Toyabe function \cite{Toyabe} and describes the muon spin depolarization in the presence of a dense network of nuclear spins or an additional electronic contribution. Here, $A_0$ denotes the maximal asymmetry, and the width of the Gaussian distribution is described by $\sigma / \gamma_{\mu}$.  \\

\textbf{Fitting of the TF-$\mu$SR data}:
The fits to the Transverse Field (TF) $\mu$SR data were performed using the following equation:
\begin{equation}
     A_{\rm TF}=A_0 \left[ \cos\left(2\pi\nu t + \frac{\pi \varphi}{180}\right) \times e^{-\frac{1}{2} (\sigma t)^2} \right]
\end{equation}
which describes the field distribution felt by the muons which stop in the sample \cite{Bastian}. Here, the $A_0$ is the maximal value of the asymmetry; the Larmor frequency at which the muon spins are oscillating is described by $\nu$; the phase shift is captured by $\varphi$ and should be near 0$^\circ$; and each component can be represented by a Gaussian distribution of width $\sigma/\gamma_{\mu}$. In the normal state, the field distribution should be fairly narrow and centered around the applied magnetic field (in this case, 30~mT; see Fig. S 1d). In the superconducting state, however, the distribution should experience broadening due to the increased distribution of local magnetic fields felt by the implanted muons. This required a two-component Gaussian fit of the data (see Fig. S 1a\&b), using the following equation:
\begin{equation}
A_{\rm TF, SC}=
 \sum_{j=1,2}A_j\left[ \cos\left(2\pi\nu_j t + \frac{\pi \varphi_j}{180}\right) \times e^{-\frac{1}{2} (\sigma_j t)^2} \right]
\end{equation}
where $j=1$ denotes the contribution to the first Gaussian distribution and $j=2$ denotes the second Gaussian component \cite{Bastian}. Examples of single- and two-component Gaussian fits to the data may be seen in the Supplementary Information in Fig. S 1a-d and in the main body of the text, in Figure 2a$\&$b, in both the time and frequency domains. When performing the fits to the data in the time domain, the asymmetry of each Gaussian component was allowed to vary, as well as the Larmor frequency and relaxation rate. In this way, the two Gaussian distributions combined may accurately capture the typical field distribution for a well-arranged triangular/hexagonal vortex lattice in a Type II superconductor. When we tried fitting with an additional exponential damping rate, this additional contribution refined to 0, so an additional exponential relaxation rate in the superconducting state was excluded from our analysis.
\\
\begin{figure*}[t!]
\centering
\includegraphics[width=0.85\linewidth]{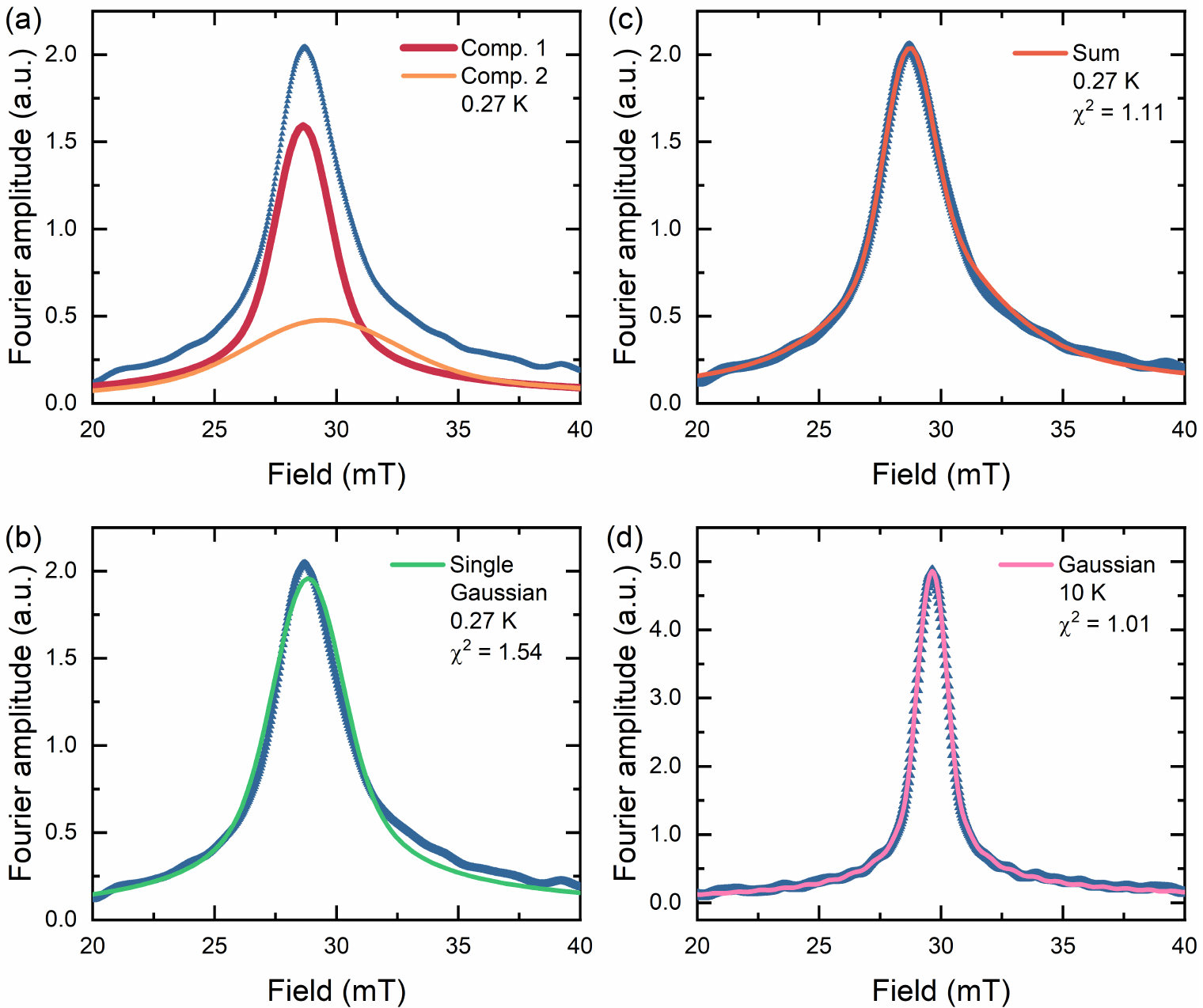}
\vspace{-0.3cm}
\caption{ (Color online)  
(a) Decomposed two Gaussian component fit compared to the Fourier transform of the ${\mu}$SR asymmetry spectra of CeRu$_{2}$ in the superconducting state at 0.27~K (blue points). (b) The sum of the two Gaussian components in panel (a) shows excellent agreement with the observed lineshape. (c) Fit to the data in the superconducting state with a single Gaussian component, showing poor agreement evidenced by the higher $\chi^2$ value. (d) Fit with a single Gaussian in the normal state, showing excellent agreement. }
\label{fig1}
\end{figure*}

\textbf{Analysis of the two-component Gaussian TF-$\mu$SR data}:
In the superconducting state, the muon ensemble experiences an increased depolarization rate due to the increased spread of the field probability distribution, which has a typical shape. In order to fit this increased, asymmetrical field distribution, we used a combination of two Gaussian components in the superconducting state. The properties of these independent Gaussian distributions (ie. field, depolarization rate $\sigma$) must be combined to calculate a single central field, $B_{\rm central}$, and overall $\sigma$, as follows:
\begin{equation}
B_{\rm central}=
\frac{A_1B_1}{A_1 + A_2}+\frac{A_2B_2}{A_1+A_2}
\end{equation}
where $A_{1,2}$ are the asymmetries of the first and second Gaussian components, respectively, and $B_{1,2}$ are the magnetic fields sensed by the muon ensemble fitted from the first and second Gaussian components, respectively. For the depolarization rate $\sigma$, the following expression is used:

\begin{widetext}
\begin{equation}
\sigma =
2\pi\gamma_{\mu}\sqrt{\frac{A_1B_1}{A_1 + A_2} \left[ \frac{\sigma_{\rm 1}^2}{4\pi^2\gamma_{\mu}^2} + \left( B_1 - B_{\rm central} \right)^2 \right] +\frac{A_2B_2}{A_1+A_2} \left[ \frac{\sigma_{\rm 2}^2}{4\pi^2\gamma_{\mu}^2} + \left( B_2 - B_{\rm central} \right)^2 \right] }
\end{equation}
\end{widetext}
where $\gamma_{\mu}$ is the gyromagnetic ratio of the muon (=~135.5~$\frac{\rm MHz}{\rm T}$) and $\sigma_{1,2}$ is the muon depolarization rate from the first and second Gaussian components, respectively.\\  




\textbf{Analysis of ${\lambda}(T)$}: ${\lambda}$($T$) was calculated within the local (London) approximation (${\lambda}$ ${\gg}$ ${\xi}$) by the following expression \cite{Bastian,Tinkham}:
\begin{equation}
\frac{\lambda^{-2}(T,\Delta_{0})}{\lambda^{-2}(0,\Delta_{0})}=
1+\frac{1}{\pi}\int_{0}^{2\pi}\int_{\Delta(_{T,\phi})}^{\infty}(\frac{\partial f}{\partial E})\frac{EdEd\phi}{\sqrt{E^2-\Delta_i(T,\phi)^2}},
\end{equation}
where $f=[1+\exp(E/k_{\rm B}T)]^{-1}$ is the Fermi function, ${\phi}$ is the angle along the Fermi surface, and ${\Delta}_{i}(T,{\phi})={\Delta}_{0}{\Gamma}(T/T_{\rm c})g({\phi}$)
(${\Delta}_{0}$ is the maximum gap value at $T=0$). The temperature dependence of the gap is approximated by the expression 
${\Gamma}(T/T_{\rm c})=\tanh{\{}1.82[1.018(T_{\rm c}/T-1)]^{0.51}{\}}$,\cite{carrington} 
while $g({\phi}$) describes the angular dependence of the gap and it is replaced by 1 for an $s$-wave gap, 
${\mid}1~+~a\cos(4{\theta}){\mid}$ for the anisotropic $s$-wave (where $a$ describes the anisotropy of the gap, with extremal values $a$~=~0 for a fully isotropic $s$-wave gap and $a$~=~1 for a strongly anisotropic gap),
and ${\mid}\cos(2{\theta}){\mid}$ for a $d$-wave gap.\\

\bibliographystyle{prsty}